\documentclass[aps,prc,showpacs,showkeys,twoside,twocolumn,byrevtex,floatfix]{revtex4-2}

\usepackage[dvips]{graphicx} 
\usepackage{color}           

\usepackage{amsmath,amssymb,amsfonts}
\usepackage{bm}
\usepackage{mathtools}
\newcommand{\nc}{\newcommand}           
\nc{\vc}[1]     {\mbox{\boldmath $#1$}} 
\nc{\mapleft}[1]{                       
 \smash{\mathop{                      %
  \hbox to 0.90cm{\rightarrowfill} }\limits_{#1}}}

\nc{\beq}     {\begin{eqnarray}}
\nc{\eeq}    {\end{eqnarray}}
\nc{\bra}       {\langle}               
\nc{\ket}       {\rangle}               
\nc{\bras}[1]   {\langle#1|}            
\nc{\kets}[1]   {|#1\rangle}            
\nc{\del}       {\partial}              
\nc{\wtil}      {\widetilde}            
\nc{\hO}        {\hat{O}}           
\nc{\EV}[1]     {\langle #1 \rangle} 

\newcommand{\lw}[1]{\smash{\lower1.75ex\hbox{#1}}}

\nc{\mydraft}	{\setlength{\topmargin}{-1.5cm}}
\mydraft
\begin{document}

\title{Resonances and scattering in microscopic cluster models with the complex-scaled generator coordinate method}

\author{Takayuki Myo\footnote{takayuki.myo@oit.ac.jp}}
\affiliation{General Education, Faculty of Engineering, Osaka Institute of Technology, Osaka, Osaka 535-8585, Japan}
\affiliation{Research Center for Nuclear Physics (RCNP), Osaka University, Ibaraki, Osaka 567-0047, Japan}

\author{Hiroki Takemoto\footnote{hiroki.takemoto@ompu.ac.jp}}
\affiliation{Faculty of Pharmacy, Osaka Medical and Pharmaceutical University, Takatsuki, Osaka 569-1094, Japan} 
\affiliation{Research Center for Nuclear Physics (RCNP), Osaka University, Ibaraki, Osaka 567-0047, Japan}

\date{\today}

\begin{abstract}%
  The generator coordinate method of a microscopic cluster model is developed to treat the resonance and scattering of nuclear clusters with complex scaling.
  We consistently derive the formulation of the complex scaling for the microscopic cluster model, in which only the relative motions between clusters are transformed in the generator coordinate wave function.
  We also reveal the applicability of this method to the cluster wave function.
  Furthermore, we demonstrate this framework in the 2$\alpha$ system of $^8$Be and obtain the solutions of resonance and non-resonant continuum states.
  Using these solutions, we calculate the level density, which brings the phase shifts of the cluster-cluster scattering.
  This work becomes the foundation in the description of the multicluster scattering states of nuclei in a microscopic framework with complex scaling. 
\end{abstract}

\maketitle

\section{Introduction}\label{sec:intro}
Clustering is a general phenomenon in nuclei \cite{ikeda68,horiuchi12,freer18},
in which some nucleons in nuclei are localized spatially and form a cluster such as an $\alpha$ particle.
A typical case is the $^8$Be nucleus, which is unbound and decays into two $\alpha$ particles.
By adding one $\alpha$ particle to $^8$Be, the $0^+_2$ state in $^{12}$C (Hoyle state) is known to have a 3$\alpha$ structure
and this state is the resonance located just above the threshold energy of the 3$\alpha$ emission.
Nuclear cluster states are often observed as resonances near and above the threshold energy of the $\alpha$-particle emissions \cite{horiuchi12}.

The resonance can be defined as a decaying state by imposing the boundary condition of the outgoing wave \cite{gamow28,siegert39}.
Under this condition, the resonance has a complex energy eigenvalue and its radial wave function shows diverging behavior in the asymptotic region.
There are several methods of treating resonances with complex energies.
The complex scaling method is one of them, used to describe the resonances in various kinds of quantum systems \cite{ho83,moiseyev98,kruppa88,lazauskas05,aoyama06,myo14,dote18,myo20}.
In this method, the boundary condition of resonance is transformed into the damping behavior, the same as that of the bound state.
Owing to this property, the complex scaling is a promising way to describe the resonance of nuclear many-body systems.
By using the Green's function with complex scaling, many-body scattering states can be described \cite{myo14,myo98,suzuki05,kruppa07},
and this framework is useful to investigate many-particle emissions of stable and unstable nuclei
such as photodisintegrations and Coulomb breakup reactions \cite{myo14,odsuren15,matsumoto10}.

In nuclear cluster models, the resonating group method (RGM) and the orthogonality condition model (OCM) have been combined with the complex scaling \cite{kruppa88,myo14,pichler97,arai06,kruppa90,kurokawa07}.
In these methods, the relative motions between clusters are solved directly and this property is compatible with the complex scaling to describe the resonances in the dynamics of the relative motions.

The Bloch--Brink (BB) $\alpha$ cluster model is also a microscopic cluster model \cite{brink66},
and this model is combined with the generator coordinate method (GCM) as BB+GCM, which is flexible to treat various nuclear systems with scattering \cite{horiuchi12,descouvemont10}. 
There are works to calculate resonances in GCM introducing the absorbing boundary potential \cite{ito05,ito06}.
As an extension of the BB cluster model, the antisymmetrized molecular dynamics (AMD) has been developed \cite{kanada03},
and recently the analytical continuation of the coupling constant (ACCC) was applied to AMD+GCM \cite{takatsu23}
to discuss the resonances in unstable nuclei from the extrapolation of the bound-state solutions.

It is shown that the GCM for the cluster model is equivalent to RGM \cite{horiuchi70},
however, in GCM the relative motion is implicit, which makes it difficult to apply the complex scaling to GCM.
Recently, Zhang et al. described the resonances of the 2$\alpha$ system of $^8$Be in BB+GCM with complex scaling \cite{zhang22}.
They transformed the parameters of the generator coordinates in each cluster to the complex-scaled ones instead of the transformation of the Hamiltonian
and obtained the complex energy of resonances.
This transformation reasonably works, but is not straightforward in its derivation.
One needs to verify this transformation and clarify the condition of this transformation on the cluster wave function.
These points are essential to develop the resonance treatment of nuclei in the GCM approach, and this is the main purpose of the present paper.

The complex scaling method can provide the level density, the scattering phase shifts,
  and the various strength functions via the Green's function \cite{myo14,suzuki05,kruppa07,odsuren14,odsuren21}.
This property of the complex scaling method enables us to treat many-body scattering states consisting of multiple cluster of nuclei,
which so far have been investigated in OCM using the intercluster potential \cite{myo14,yamamoto09};
however not yet done in a microscopic way as in RGM and GCM.
It is important to obtain the level density, phase shifts, and strength functions with complex scaling in the microscopic description of nuclear scattering.

In this paper, we provide a consistent derivation of the complex scaling for BB+GCM and clarify its physical meaning in relation to the relative motion between clusters.
We also reveal the applicability of this method to the cluster wave function.
Furthermore, we confirm the reliability of the method by calculating the level density of the 2$\alpha$ system of $^8$Be,
in which we use the complex-scaled solutions of the resonant and nonresonant continuum states obtained in GCM.
We finally evaluate the phase shift of the $\alpha$--$\alpha$ scattering in a microscopic way.
The present work becomes the foundation to investigate the multicluster scattering phenomena in the GCM approach with complex scaling.

In Sec.~\ref{sec:method}, we derive the formulation to apply the complex scaling method to the BB $\alpha$-cluster model in the GCM calculations.
In Sec.~\ref{sec:result}, we discuss the resonances and scattering of the 2$\alpha$ system of $^8$Be.
In Sec.~\ref{sec:summary}, we summarize this work.

\section{Theoretical methods}\label{sec:method}

\subsection{Complex scaling method}\label{sec:csm}

We describe many-body resonances using the complex scaling method \cite{ho83,moiseyev98,aoyama06,myo14,myo20}.
In this method, the particle coordinate $\{\vc{r}_j\}$ and the conjugate momentum $\{\vc{p}_j\}$ in the Hamiltonian $H$ and the wave function $\Psi$ are transformed using a common scaling angle $\theta$ with an operator $U(\theta)$ as
\begin{equation}
U(\theta)~:~\vc{r}_j \to \vc{r}_j\, e^{ i\theta},\qquad
\vc{p}_j \to \vc{p}_j\, e^{-i\theta} ,
\label{eq:CSM}
\end{equation}
where $j$ is the index of the degrees of freedom in the system and $\theta$ is a real positive number. 
The complex-scaled Hamiltonian $H^\theta$, wave function $\Psi^\theta$, and total energy $E^\theta$ are given as
\begin{equation}
\begin{split}
  H^\theta
  &=     U(\theta)H U^{-1}(\theta) ,\quad
  \Psi^\theta 
  =   U(\theta)\Psi = \sum_n C_n^\theta \Phi_n ,
  \\
  E^\theta &= \frac{ \langle \tilde \Psi^\theta|H^\theta|\Psi^\theta \rangle}{ \langle \tilde \Psi^\theta|\Psi^\theta \rangle} .
\end{split}
\end{equation}
The wave function $\Psi^\theta$ is expanded with the basis functions $\{\Phi_n\}$ with the index $n$,
and $\{C_n^\theta\}$ are the unknown coefficients to be determined.
From the variational principle for the energy, $\delta E^\theta=0$,
one solves the eigenvalue problem of the complex-scaled Hamiltonian matrix and obtains $E^\theta$ and $\{C_n^\theta\}$. 
The wave function $\tilde \Psi^\theta$ is the biorthogonal state of $\Psi^\theta$ \cite{berggren68}. 
One does not take the complex conjugate in the radial part of the bra state in the matrix elements.

The energy eigenvalues $E^\theta$ are obtained for bound, resonant, and continuum states in the complex energy plane according to the so-called ABC theorem \cite{ABC}.
The continuum states start from the threshold energies of the cluster emissions and are obtained with the complex energies on the lines rotated down by $2\theta$ from the real energy axis.
The energy eigenvalues of the bound and resonant states are independent of $\theta$ in principle.
The resonance has a complex energy of $E_{\rm R}=E_r-i\Gamma/2$, where $E_r$ is a resonance energy and $\Gamma$ is a decay width.
For the resonance wave function, its asymptotic behavior becomes the damping form if $2\theta > |\arg(E_{\rm R})|$ \cite{ABC}.
In the numerical calculation, one can identify the resonance in the complex energy plane
from the stationary property of $E_{\rm R}$ with respect to $\theta$ \cite{ho83,moiseyev98,aoyama06}.
In the present calculation, the continuum states are discretized in the energy eigenvalues due to the finite number of the basis states.

\subsection{Bloch--Brink $\alpha$ cluster model}\label{sec:brink}

We explain the microscopic BB $\alpha$ cluster model for the $N_\alpha\alpha$ system \cite{brink66},
where $N_\alpha$ is the number of the $\alpha$ particles and the mass number $A=4N_\alpha$.
The total wave function $\Phi_{\rm BB}(\nu)$ is a single Slater determinant of $A$-nucleons and is given as
\begin{align}
  \phi_{\gamma}(\nu,\bm{r},\bm{R})&=\left(\frac{2\nu}{\pi}\right)^{3/4}\!\! e^{-\nu(\bm{r}-\bm{R})^2} \chi_{\gamma}~,
  \label{eq:Gauss}
  \\
  \Phi_\alpha (\nu, \bm{R}) &= {\cal A} \left\{ \prod_{\gamma=1}^4 \phi_{\gamma}(\nu,\bm{r}_{\gamma},\bm{R}) \right\}~,
  \label{eq:alpha}
  \\
  \Phi_{\rm BB}(\nu) &= {\cal A}' \left\{ \prod_{i=1}^{N_\alpha} \Phi_{\alpha_i} (\nu, \bm{R}_i) \right\}~.
  \label{eq:BB}
\end{align}
The single-nucleon wave function $\phi_\gamma(\nu,\bm{r},\bm{R})$ has a Gaussian wave packet with a common range parameter $\nu$ for all nucleons
and the centroid position $\bm{R}$, which is a real number.
The spin and isospin part is $\chi_{\gamma}$ for $p_\uparrow$, $p_\downarrow$, $n_\uparrow$, and $n_\downarrow$ with an index $\gamma$.
The $\alpha$ cluster wave function is $\Phi_\alpha (\nu, \bm{R})$ with a common $\bm{R}$, being the generator coordinate of each $\alpha$ cluster.
This wave function is the $0s$ configuration.
The BB wave function $\Phi_{\rm BB}(\nu)$ for the $N_\alpha\alpha$ system has a set of $\{\bm{R}_i\}$ with $i=1,\ldots,N_\alpha$,
the summation of which is zero: $\sum_{i=1}^{N_\alpha} \bm{R}_i = \bm{0}$.
The operator ${\cal A}$ is the antisymmetrizer for the nucleons in the $\alpha$ cluster and
${\cal A}'$ is the one for the nucleons in the different $\alpha$ clusters.

We perform the projection of the intrinsic wave function $\Phi_{\rm BB}(\nu)$
on the eigenstates of the angular momentum $J$ with quantum numbers of $M$ and $K$, and also the parity ($\pm$). 
\begin{equation}
\Phi^{J^\pm}_{MK}(\nu)
= P^J_{MK}P^{\pm} \Phi_{\rm BB}(\nu),
\label{eq:projection}
\end{equation}
where $P^J_{MK}$ and $P^{\pm}$ are the projection operators.
The total energy of the $N_\alpha \alpha$ system is obtained by calculating the expectation value of the Hamiltonian $H$ given as 
\begin{equation}
    H
= \sum_{i=1}^{A} t_i - T_{\rm CM} + \sum_{i<j}^{A} v_{ij}~,
    \label{eq:Ham}
\end{equation}
where $t_i$ and $T_{\rm CM}$ are the kinetic energies of each nucleon and the center of mass (CM), respectively,
and $v_{ij}$ is the two-body nuclear and Coulomb interactions.
This form of the Hamiltonian is based on the single-particle motion and is tractable in the BB wave function.

The single BB wave function is extended to the multiconfiguration in the GCM
using various sets of the Gaussian centroids $\{\bm{R}_i\}$ in Eq.~(\ref{eq:BB}).
We employ a finite number of the BB basis states with different sets of $\{\bm{R}_i\}$ and superpose them in BB+GCM.
The total wave function $\Psi_{\rm GCM}$ is a superposition of the projected BB basis states denoted as $\Phi_n$
and the total energy $E_{\rm GCM}$ is given as
\begin{equation}
\begin{split}
   \Psi_{\rm GCM}
&=\sum_{n=1}^{N_{\rm GCM}} C_n \Phi_{n},\quad
   \\
   E_{\rm GCM}&=\frac{ \langle \tilde \Psi_{\rm GCM}|H|\Psi_{\rm GCM} \rangle}{ \langle \tilde \Psi_{\rm GCM}|\Psi_{\rm GCM} \rangle} \,,
\end{split}
\end{equation}
where we use the biorthogonal state in the bra state to apply the complex scaling.
From the variational principle for $E_{\rm GCM}$, we solve the Hill-Wheeler equation,
which results in a generalized eigenvalue problem of the Hamiltonian matrix.

\subsection{Complex-scaled generator coordinate method}\label{sec:cs-gcm}

We explain the application of the complex scaling method to the GCM of the multi-$\alpha$-cluster system.
Zhang et al. introduced the transformation of the generator coordinates of clusters: $\{\bm{R}_i\}\to\{\bm{R}_ie^{i\theta}\}$ in Eq. (\ref{eq:BB}) \cite{zhang22}.
We derive that this transformation is equivalent to the complex scaling of only the relative motions between clusters
and clarify the condition of this transformation on applying it to the GCM calculations.

In the nuclear cluster models of RGM and OCM,
the relative wave function between clusters is explicit and directly transformed with the complex scaling,
while the internal wave functions of clusters are not transformed.
On the other hand, the BB basis state is based on the single-particle picture in the Slater determinant
and the relative wave function does not appear explicitly.
Hence, it is not straightforward to apply the complex scaling to the BB basis states,
and we show here the one treatment to solve this problem. For this purpose, we discuss the 2$\alpha$ system of $^8$Be.

In the complex scaling method, we demand that the $\alpha$ cluster state is fixed as the $0s$ configuration
and transform only the relative motion between two $\alpha$ clusters, which can produce the resonances.
In the 2$\alpha$ system, the Hamiltonian $H$ in Eq.~(\ref{eq:Ham}) can be written with the internal $\alpha$ cluster part $H_\alpha$
and the relative motion part $H_{\rm rel}$ as
\begin{equation}
\begin{split}
  H &= H_{\alpha_1} + H_{\alpha_2} + H_{\rm rel},
  \\
  H_\alpha  &= \sum_{i=1}^4 t_i - T_{{\rm CM},\alpha} + \sum_{i<j}^4 v_{ij},
\end{split}
\end{equation}
where $T_{{\rm CM},\alpha}$ is the kinetic energy operator for CM of the $\alpha$ cluster.
We first define the complex-scaled Hamiltonian $\bar H ^\theta$ for the transformation of only the relative motion as
\begin{equation}
\begin{split}
  \bar H^\theta &= H_{\alpha_1} + H_{\alpha_2} + H^\theta_{\rm rel}~,
  \\
  H^\theta_{\rm rel} &= U_{\rm rel}(\theta) H_{\rm rel} U_{\rm rel}^{-1}(\theta)~,
\end{split}
\label{eq:cs-ham}
\end{equation}
where $U_{\rm rel}(\theta)$ is the operator to transform $H_{\rm rel}$.
In the BB basis states, it is difficult to calculate the matrix elements of $\bar H^\theta$ in the form of Eq.~(\ref{eq:cs-ham}).
Hence we introduce the inverse transformation of the complex scaling for $H_{\alpha_1}$ and $H_{\alpha_2}$ as
\begin{equation}
\begin{split}
  \bar H^\theta
  &= U^{-1}_{\alpha_1}(\theta) H^\theta_{\alpha_1} U_{\alpha_1}(\theta)  + U^{-1}_{\alpha_2}(\theta)  H^\theta_{\alpha_2} U_{\alpha_2}(\theta) + H_{\rm rel}^\theta~, 
  \\
  H^\theta_{\alpha_i} &= U_{\alpha_i}(\theta) H_{\alpha_i} U_{\alpha_i}^{-1}(\theta),
\end{split}
\end{equation}
where $i=1,2$. The operator $U_{\alpha}(\theta)$ transforms the internal coordinates and momenta of the $\alpha$ cluster in $H_{\alpha}$.
This operator satisfies the following commutation relations owing to the irrelevant degrees of freedom with $i\neq j$;
\begin{equation}
  \begin{split}
  \left[ U_{\alpha_i}(\theta),H_{\alpha_j}^\theta \right] &= 0,\quad
  \left[ U_{\alpha_i}(\theta),H_{\rm rel}^\theta \right]   = 0, \\
  \left[ U_{\alpha_i}(\theta),U_{\alpha_j}(\theta)\right] &= 0.
  \end{split}
\end{equation}
Using these relations, $\bar H^\theta$ can be written as
\begin{equation}
  \begin{split}
    \bar H^\theta &=  U^{-1}_{\alpha_2}(\theta) U^{-1}_{\alpha_1}(\theta)\, \Bigl\{ H^\theta_{\alpha_1} + H^\theta_{\alpha_2} + H_{\rm rel}^\theta \Bigr\}\,
    U_{\alpha_1}(\theta) U_{\alpha_2}(\theta) 
  \\
  &=  U^{-1}_{\alpha_2}(\theta) U^{-1}_{\alpha_1}(\theta) \, H^\theta \, U_{\alpha_1}(\theta) U_{\alpha_2}(\theta).
  \label{eq:ham_cs1}
  \end{split}
\end{equation}
Here $H^\theta$ is the complex-scaled Hamiltonian transformed from $H$ in Eq.~(\ref{eq:Ham}) using $U_{\rm all}(\theta)$ for all degrees of freedom;
\begin{equation}
  \begin{split}
  U_{\rm all}(\theta)&=U_{\alpha_1}(\theta) U_{\alpha_2}(\theta) U_{\rm rel}(\theta)~,
  \\
  H^\theta &= U_{\rm all}(\theta) H\, U_{\rm all}^{-1}(\theta)~.
  \end{split}
\end{equation}
The matrix elements of $H^\theta$ are calculable in the BB basis states,
because all degrees of freedom are commonly transformed with $\theta$.

Next, we operate $U_{\alpha}(\theta)$ in Eq. (\ref{eq:ham_cs1}) to the internal wave function of the $\alpha$ cluster.
Here, we omit the spin-isospin part for simplicity.
The $\alpha$ cluster wave function in Eq.~(\ref{eq:alpha}) can be decomposed into the internal and CM parts and the internal part $\Phi^{\rm int}_\alpha (\nu)$ is given
with the Jacobi coordinates $\{\tilde{\bm{r}}_k\}$ and the corresponding range parameters $\{\tilde{\nu}_k\}$ with $k=1,2,$ and 3 \cite{horiuchi70},
defined as
\begin{equation}
  \begin{split}
  \Phi_\alpha^{\rm int}(\nu) &= {\cal A} \left\{ \prod_{k=1}^3 \left( \frac{2\tilde{\nu}_k}{\pi} \right)^{3/4}\!\! e^{-\tilde{\nu}_k \tilde{\bm{r}}_k^2} \right\} ,
  \\
\tilde{\bm{r}}_k &= \bm{r}_{k+1}-\frac{1}{k} \sum_{i=1}^k \bm{r}_i ,
\quad
\tilde{\nu}_k~=~\frac{k}{k+1} \nu .
  \end{split}
  \end{equation}
The operator $U_\alpha(\theta)$ acts on only the internal wave function of the $\alpha$ cluster:
\begin{equation}
  \begin{split}
U_\alpha(\theta) \Phi_\alpha^{\rm int}(\nu) 
&= 
{\cal A} \left\{ \prod_{k=1}^3 e^{3i\theta/2} \left( \frac{2\tilde{\nu}_k}{\pi} \right)^{3/4}\!\! e^{-\tilde{\nu}_k \tilde{\bm{r}}_k^2 e^{2i\theta}} \right\}
\\
&= 
  {\cal A} \left\{ \prod_{k=1}^3 \left( \frac{2\tilde{\nu}_k e^{2i\theta}}{\pi} \right)^{3/4}\!\! e^{-( \tilde{\nu}_k e^{2i\theta} ) \tilde{\bm{r}}_k^2 } \right\}
  \\
&=
\Phi_\alpha^{\rm int}(\nu e^{2i\theta})
\eqqcolon U_{\alpha,\nu}(2\theta) \Phi_\alpha^{\rm int}(\nu)~,
\label{eq:alpha_cs}
  \end{split}
  \end{equation}
where the factor $e^{3i\theta/2}$ comes from the Jacobian.
We define the operator $U_{\alpha,\nu}(2\theta)$ to transform the range parameter $\nu$ to $\nu e^{2i\theta}$ in the internal wave function of the $\alpha$ cluster.
Equation (\ref{eq:alpha_cs}) shows an important property: no other parameters are involved in the transformation.
We can write $\bar H^\theta$ in Eq.~(\ref{eq:ham_cs1}) with $U_{\alpha,\nu}(2\theta)$ supposing the application to the BB wave function as
\begin{equation}
  \bar H^\theta =  U^{-1}_{\alpha_2,\nu}(2\theta) U^{-1}_{\alpha_1,\nu}(2\theta)\,  H^\theta\, U_{\alpha_1,\nu}(2\theta) U_{\alpha_2,\nu}(2\theta) .
  \label{eq:ham_cs2}
\end{equation}
We set the BB wave function of the $2\alpha$ system with the generator coordinates $\bm{R}_1=\bm{R}$ and $\bm{R}_2=-\bm{R}$
and extract the relative and CM wave functions between $2\alpha$ \cite{horiuchi70} as
\begin{equation}
  \begin{split}
  \Phi_{\rm BB} (\nu)
  &= {\cal A}' \left\{\Phi_{\alpha_1}(\nu, \bm{R})\cdot \Phi_{\alpha_2} (\nu,-\bm{R}) \right\}
  \\
  &= {\cal A}' \left\{\Phi^{\rm int}_{\alpha_1}(\nu)\Phi^{\rm CM}_{\alpha_1} (\nu,\bm{R}) \cdot \Phi^{\rm int}_{\alpha_2} (\nu) \Phi^{\rm CM}_{\alpha_2} (\nu,-\bm{R}) \right\}
  \\
  &= {\cal A}' \left\{\Phi^{\rm int}_{\alpha_1}(\nu)\Phi^{\rm int}_{\alpha_2} (\nu) \cdot  \Phi^{\rm rel}(\nu)  \right\} \Phi^{\rm CM}_{2\alpha} (\nu)~.
  \label{eq:BB2}
  \end{split}
  \end{equation}
The relative and CM wave functions, $\Phi^{\rm rel}(\nu)$ and $\Phi^{\rm CM}_{2\alpha} (\nu)$,
are given with the relative and the CM coordinates of $\bm{r}$ and $\bm{r}_{\rm G}$, respectively, as
\begin{equation}
\begin{split}
  \Phi^{\rm rel}(\nu)
&=\left(\frac{4\nu}{\pi}\right)^{3/4}\!\! e^{-2\nu(\bm{r}-2\bm{R})^2} ,
  \\
  \Phi^{\rm CM}_{2\alpha} (\nu)
&=\left(\frac{16\nu}{\pi}\right)^{3/4}\!\! e^{-8\nu\bm{r}_{\rm G}^2} .
 \end{split}
\end{equation}
We try to apply the complex scaling to $\Phi_{\rm BB} (\nu)$ of $2\alpha$ in Eq.~(\ref{eq:BB2}), expanding the antisymmetrization as
\begin{align*}
&  U_{\alpha_1,\nu}(2\theta) U_{\alpha_2,\nu}(2\theta) \Phi_{\rm BB} (\nu)
  \\
  &=
  \Phi^{\rm int}_{\alpha_1}(\nu e^{2i\theta})\cdot \Phi^{\rm int}_{\alpha_2} (\nu e^{2i\theta}) \cdot
  \Phi^{\rm rel}(\nu)\cdot \Phi^{\rm CM}_{2\alpha} (\nu) + \cdots .
\end{align*}
In this equation, the range parameters are different among the internal, relative, and CM parts.
From this form it is difficult to calculate the matrix elements with the BB basis states.

It is noticed here that the relative wave function between clusters is unknown in the GCM calculation
and the superposition of the relative wave function $\Phi^{\rm rel}(\nu)$ in the BB basis states is the function to be determined.
This means that we can replace this part with $\Phi^{\rm rel}(\nu e^{2i\theta})$ for the basis states to be superposed.
Similarly, the CM wave function, $\Phi^{\rm CM}_{2\alpha} (\nu)$, does not affect any solutions in the GCM calculation, 
and then we can replace it with $\Phi^{\rm CM}_{2\alpha} (\nu e^{2i\theta})$.
Following these properties, the internal, relative, and CM parts of the BB wave function can have the same dependence of $\nu e^{2i\theta}$ and then we can define the transformed wave function $\Phi_{\rm BB}(\nu e^{2i\theta})$ as follows
\begin{equation}
  \begin{split}
&  U_{\alpha_1,\nu}(2\theta) U_{\alpha_2,\nu}(2\theta)\Phi_{\rm BB}(\nu)
 \\ 
  &\to {\cal A}' \left\{\Phi^{\rm int}_{\alpha_1}(\nu e^{2i\theta})\cdot \Phi^{\rm int}_{\alpha_2} (\nu e^{2i\theta}) \cdot  \Phi^{\rm rel}(\nu e^{2i\theta})  \right\}  \Phi^{\rm CM}_{2\alpha} (\nu e^{2i\theta})
  \\
  &= {\cal A}' \left\{\Phi_{\alpha_1}(\nu e^{2i\theta}, \bm{R})\cdot \Phi_{\alpha_2} (\nu e^{2i\theta},-\bm{R}) \right\}
  \\
  &= \Phi_{\rm BB}(\nu e^{2i\theta}) 
  \eqqcolon U_{\nu}(2\theta) \Phi_{\rm BB}(\nu) ,
  \end{split}
  \end{equation}
Here we can express the transformed BB wave function in the Slater determinant because all nucleons have a common range parameter $\nu e^{2i\theta}$,
and we define the operator $U_{\nu}(2\theta)$ for the total system.
Finally, we express $\bar H^\theta$ in Eq.~(\ref{eq:ham_cs2}) with $U_{\nu}(2\theta)$ as
\begin{equation}
  \bar H^\theta =  U^{-1}_{\nu}(2\theta) H^\theta U_{\nu}(2\theta) .
  \label{eq:ham_cs3}
\end{equation}
We use this complex-scaled Hamiltonian $\bar H^\theta$ in the BB+GCM calculation.
The physical meaning of $\bar H^\theta$ is that only the relative motion between clusters is transformed.
The matrix elements of $\bar H^\theta$ with the BB basis states, $\Phi_m$ and $\Phi_n$, are given as
\begin{equation}
  \begin{split}
    & \langle \tilde \Phi_m(\nu,\bm{R}) |\bar H^\theta| \Phi_n(\nu,\bm{R}') \rangle
    \\
  &= \langle \tilde \Phi_m(\nu e^{2i\theta},\bm{R}) |H^\theta| \Phi_n(\nu e^{2i\theta},\bm{R}') \rangle
  \\
  &= \langle \tilde \Phi_m(\nu e^{2i\theta},\bm{R}) |U_{\rm all}(\theta)\, H\, U_{\rm all}^{-1}(\theta)|\Phi_n(\nu e^{2i\theta},\bm{R}') \rangle,
  \end{split}
  \end{equation}
where we write the two arguments in the BB basis states to show the operation of the complex scaling. The vectors $\bm{R}$ and $\bm{R}'$ stand for the sets of generator coordinates $\{{\bm R}_i\}$ and $\{{\bm R}_i'\}$ with $i=1,2$ for $2\alpha$, respectively.
We operate $U_{\rm all}^{-1}(\theta)$ to the single-nucleon wave functions $\{\phi\}$ in Eq.~(\ref{eq:Gauss})
having a range parameter $\nu e^{2i\theta}$ and the generator coordinate $\bm{R}_i$ in the BB basis states as
\begin{equation}
  \begin{split}
  &~ U_{\rm all}^{-1}(\theta) \phi(\nu e^{2i\theta},\bm{r},\bm{R}_i)
  \\
  &= e^{-3i\theta/2} \left(\frac{2\nu e^{2i\theta}}{\pi}\right)^{3/4}\!\! e^{-\nu e^{2i\theta}(\bm{r}e^{-i\theta}-\bm{R}_i)^2} 
  \\
 &= \left(\frac{2\nu}{\pi}\right)^{3/4}\!\! e^{-\nu (\bm{r}-\bm{R}_i e^{i\theta})^2} 
  = \phi(\nu,\bm{r},\bm{R}_i e^{i\theta})~.
  \end{split}
  \end{equation}
It is found that the generator coordinate is transformed into $\bm{R}_i e^{i\theta}$, while the range parameter returns to $\nu$. 
This transformation is used in Ref. \cite{zhang22}. We give the complex-scaled matrix elements of
the Hamiltonian, $\bar H^\theta_{mn}$, and norm, $\bar N^\theta_{mn}$, as
\begin{equation}
  \begin{split}
  \bar H^\theta_{mn}
  &= \langle \tilde \Phi_m(\nu,\bm{R}e^{i\theta}) |H|\Phi_n(\nu,\bm{R}'e^{i\theta}) \rangle,
  \\
  \bar N^\theta_{mn}
  &= \langle \tilde \Phi_m(\nu,\bm{R}e^{i\theta}) |\Phi_n(\nu,\bm{R}'e^{i\theta}) \rangle.
  \end{split}
  \end{equation}
We omit the notation of the $J^\pm$ projections for simplicity.
It is noted that the norm matrix is transformed and then is not positive definite.
The eigenvalue problem to get the total energy $E_{\rm GCM}^\theta$ is given as
\begin{equation}
   \sum_{n=1}^{N_{\rm GCM}} \left( \bar H_{mn}^\theta - E_{\rm GCM}^\theta \bar N_{mn}^\theta \right) C_n^\theta = 0.
   \label{eq:eigen}
\end{equation}
The energy $E_{\rm GCM}^\theta$ becomes complex number because the relative motion between clusters is complex-scaled,
while the internal energies of clusters retain to be real numbers without the complex scaling.

The present framework can be extended to multi-$\alpha$-cluster systems such as $3\alpha$
to treat many-body resonances in the complex scaling method. 
The addition of valence nucleons is also available, such as $\alpha+\alpha+n$.
The condition of the framework is given in Eq.~(\ref{eq:alpha_cs}) for the internal wave function of a cluster:
the complex scaling to the physical coordinates is equivalent to the complex scaling to the range parameter $\nu$. 
This means that the internal wave function of a cluster cannot have the generator coordinates, and then the harmonic oscillator shell model wave function is applicable as in $^{16}$O and $^{40}$Ca.
This is because the harmonic oscillator basis function is a function of $\sqrt{\nu}r$ in the radial part with $\nu=m\omega/(2\hbar)$ and satisfies the condition of Eq.~(\ref{eq:alpha_cs}). 
The configuration mixing is also available with the separation of the CM motion of cluster.
Using the Green's function with complex scaling, one can investigate the multicluster scattering states
involving the effect of resonances under the correct boundary condition \cite{myo14}.

\subsection{Level density}\label{sec:level}

In the complex scaling method, the completeness relation is expressed in terms of the solutions of the bound (B), resonant (R),
and non-resonant continuum (C) states \cite{myo98,berggren68} given as
\begin{equation}
  1      = \sum_{n \in {\rm B,R,C}} \kets{\Psi_n^\theta}\bras{\tilde \Psi_n^\theta},
          \label{eq:ECR}
\end{equation}
where $n$ is the state index. Using the complex-scaled solutions of $\{\Psi_n^\theta,\tilde \Psi_n^\theta\}$ and the energy eigenvalues $\{E_n^\theta\}$,
one can introduce the complex-scaled Green's function ${\cal G}^\theta(E)$: 
\begin{equation}
	{\cal G}^\theta(E)
=	\frac{ 1 }{ E-\bar H^\theta }
=	\sum_n \frac{|\Psi^\theta_n\rangle \langle \tilde{\Psi}^\theta_n|}{E-E_n^\theta} . 
	\label{eq:green1}
\end{equation}
We apply the complex scaling to the level density $\rho(E)=\sum_{n}\delta(E-E_n)$ and use ${\cal G}^\theta(E)$ \cite{suzuki05,odsuren14,odsuren21}.
The complex-scaled level density $\rho^\theta(E)$ is given as
\begin{equation}
        \rho^\theta(E)
=     -\frac1{\pi}\ {\rm Im}\left\{{\rm Tr}\, {\cal G}^\theta(E) \right\} 
=     -\frac1{\pi}\ \sum_n {\rm Im}\left( \frac{1}{E-E_n^\theta} \right). 
        \label{eq:LD}
\end{equation}
We also consider the asymptotic Hamiltonian $\bar H_0^\theta$, omitting the finite-range interaction between clusters
from the full Hamiltonian $\bar H^\theta$.
The energy eigenvalues are $\{E_{0,n}^\theta\}$ and the asymptotic level density $\rho_0^\theta(E)$ is given as
\begin{equation}
        \rho_0^\theta(E)
=       -\frac1{\pi}\ \sum_n {\rm Im}\Biggl( \frac{1}{E-E_{0,n}^\theta} \Biggr) . 
        \label{eq:LD0}
\end{equation}
The difference between $\rho^\theta(E)$ and $\rho^\theta_0(E)$ is the so-called continuum level density $\Delta(E)$
representing the effect of the interaction in the level density. It is shown that $\Delta(E)$ is obtained independently of $\theta$ \cite{suzuki05}.
It is known that $\Delta(E)$ has a relation to the scattering matrix $S(E)$ \cite{levine69,kruppa98} as 
\begin{equation}
  \Delta(E) =\rho^\theta(E) - \rho_0^\theta(E)
  =\frac1{2\pi} {\rm Im} \frac{{\rm d}}{{\rm d} E} {\rm ln}\bigl\{{\rm det}\, S(E)\bigr\} .
\end{equation}
In the single channel problem, $\Delta(E)$ gives the derivative of the phase shift as
\begin{equation}
  \Delta(E) = \frac1\pi \frac{{\rm d}\delta(E)}{{\rm d} E} .
  \label{eq:CLD}
\end{equation}
We obtain the phase shift from the integral of $\Delta (E)$ as
\begin{equation}
  \delta(E) = \pi \int_{-\infty}^E \Delta (E') {\rm d}E'.
  \label{eq:phase}
\end{equation}
Using the energy eigenvalues in the complex scaling, one can evaluate the phase shift of the cluster-cluster scattering.
Numerically, we check the stationary property of the solutions with respect to the scaling angle $\theta$,
  because of the finite number of the basis states.
This framework of the level density can be applied to the many-body scattering states straightforwardly as in the $3\alpha$ system \cite{myo14}.

We define the asymptotic Hamiltonian $H_0$ of the $2\alpha$ system \cite{kamimura77} before applying the complex scaling:
\begin{equation}
    H_0
=   \sum_{i=1}^{8} t_i - T_{\rm CM} + \sum_{i<j \in \alpha_1}\!\!\! v_{ij} + \sum_{i<j \in \alpha_2}\!\!\! v_{ij} +  \frac{Z_1 Z_2 e^2}{r_{\alpha_1 \alpha_2}} .
    \label{eq:Ham_aysm}
\end{equation}
We include the nuclear and Coulomb interactions in each $\alpha$ cluster, but omit the nuclear interaction for the intercluster part.
The intercluster Coulomb interaction is replaced with the point type, where $Z_i$ is the charge number of $\alpha_i$ and $r_{\alpha_1 \alpha_2}=| \bm{r}_{\alpha_1} - \bm{r}_{\alpha_2} |$ is the intercluster distance.
The coordinate $\bm{r}_{\alpha}$ represents the CM position of the $\alpha$ cluster.
We also define the asymptotic BB wave function $\Phi_{\rm BB,0}$ of $2\alpha$ omitting the antisymmetrization between the nucleons in the different $\alpha$ clusters \cite{kamimura77} as
\begin{equation}
  \Phi_{\rm BB,0} = \Phi_{\alpha_1} (\nu,\bm{R}_1) \cdot \Phi_{\alpha_2} (\nu,\bm{R}_2) .
\end{equation}
We calculate the complex-scaled matrix elements of $\bar H_0^\theta$ and norm with $\Phi_{\rm BB,0}$ and solve the eigenvalue problem
to get the energy $\{E_{0,n}^\theta\}$.

\section{Results}\label{sec:result}

\subsection{$\alpha$-$\alpha$ resonances}
In this study, we treat the $2\alpha$ system of $^8$Be and discuss the $\alpha$-$\alpha$ resonances in the GCM calculation.
We use the effective nucleon-nucleon interaction of the Volkov No.2 central force with Majorana parameter $M=0.6$ \cite{arai06,volkov65}
and the point Coulomb force for protons.
In the BB wave function, we use the range parameter $\nu=0.264$ fm$^{-2}$ in Eq. (\ref{eq:Gauss}), 
which minimizes the energy of the $\alpha$ particle, $E_\alpha=-27.96$ MeV. The matter radius of the $\alpha$ particle is 1.46 fm.
For generator coordinates of the $2\alpha$ system, we employ 30 basis states with the mean relative distance between two $\alpha$ clusters
from 2/3 to 20 fm at equal intervals, which are sufficient to converge the solutions.
In particular, the basis states with long intercluster distances tend to contribute to
the continuum level density and phase shift near the $\alpha+\alpha$ threshold energy.

\begin{figure*}[t]  
\centering
\includegraphics[width=7.5cm,clip]{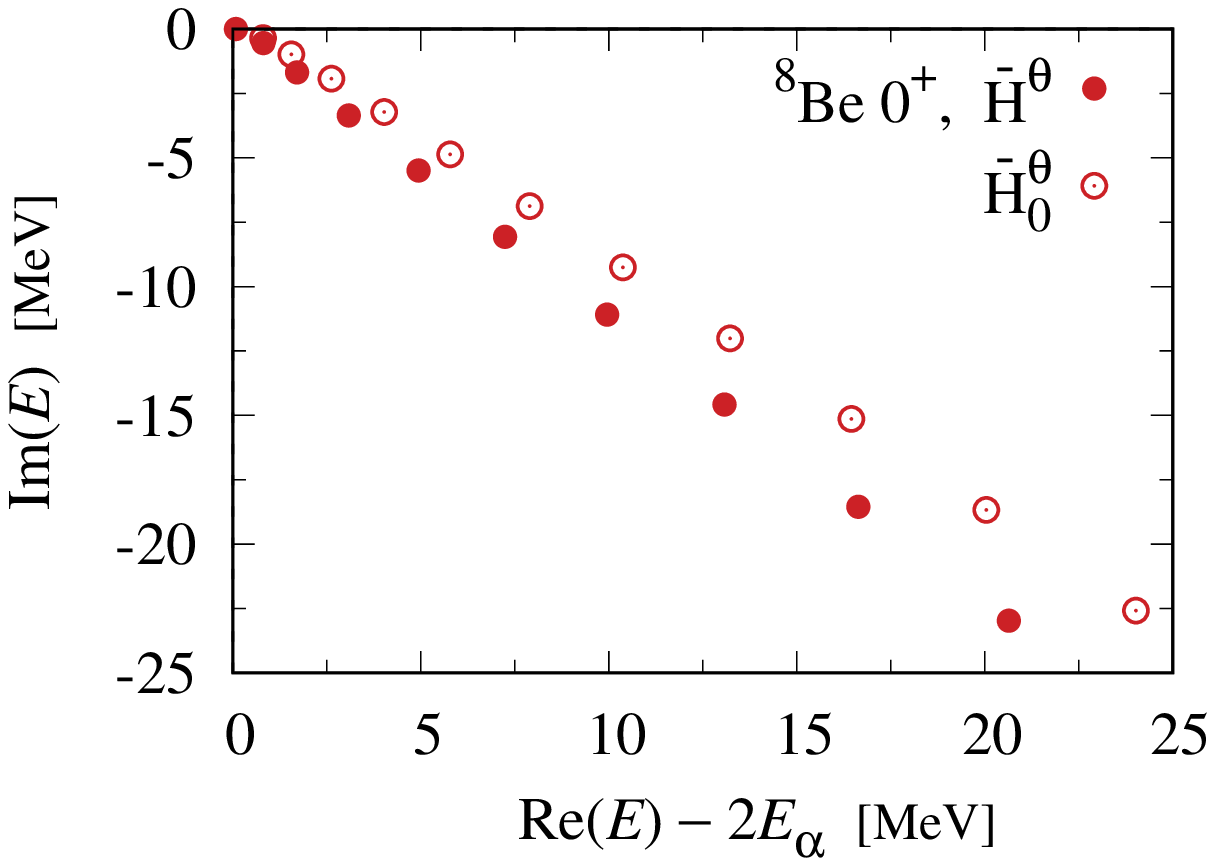}~~~~~~
\includegraphics[width=7.5cm,clip]{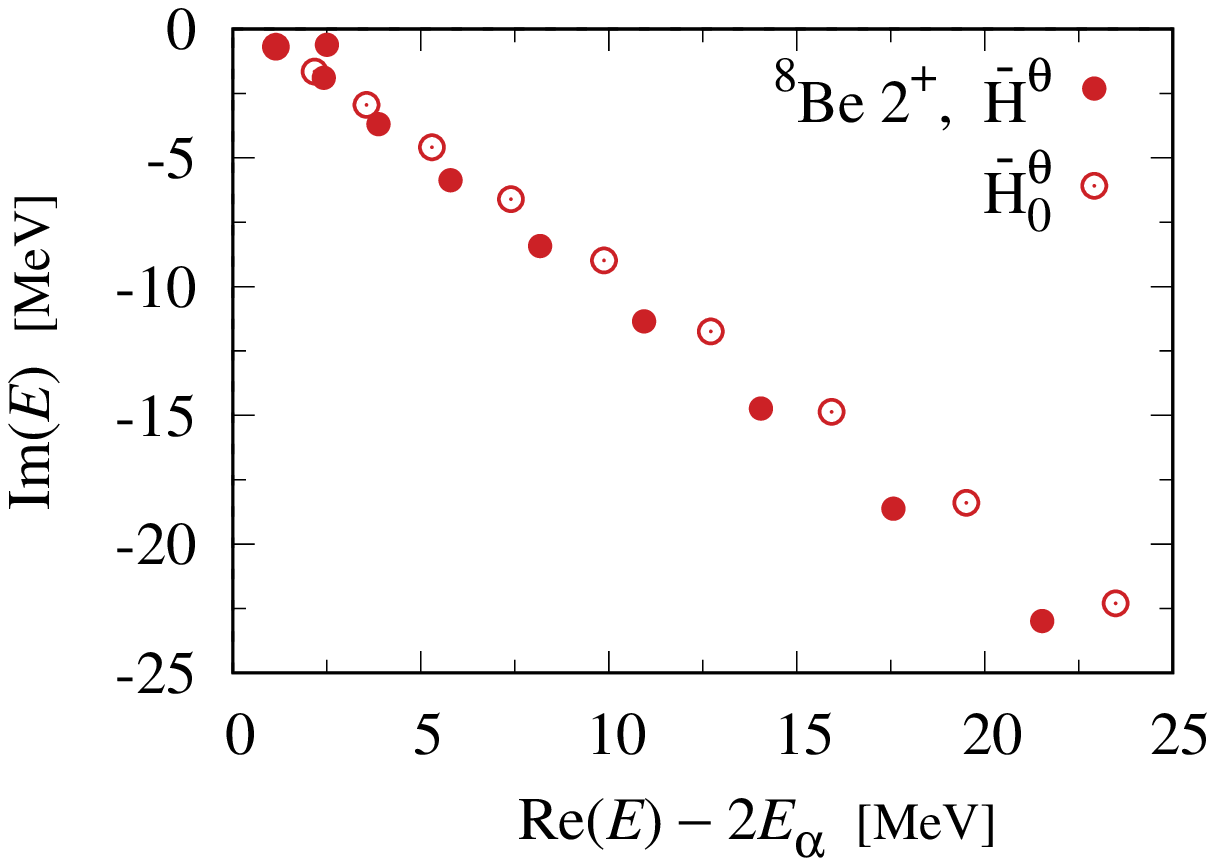}\\[0.2cm]
\includegraphics[width=7.5cm,clip]{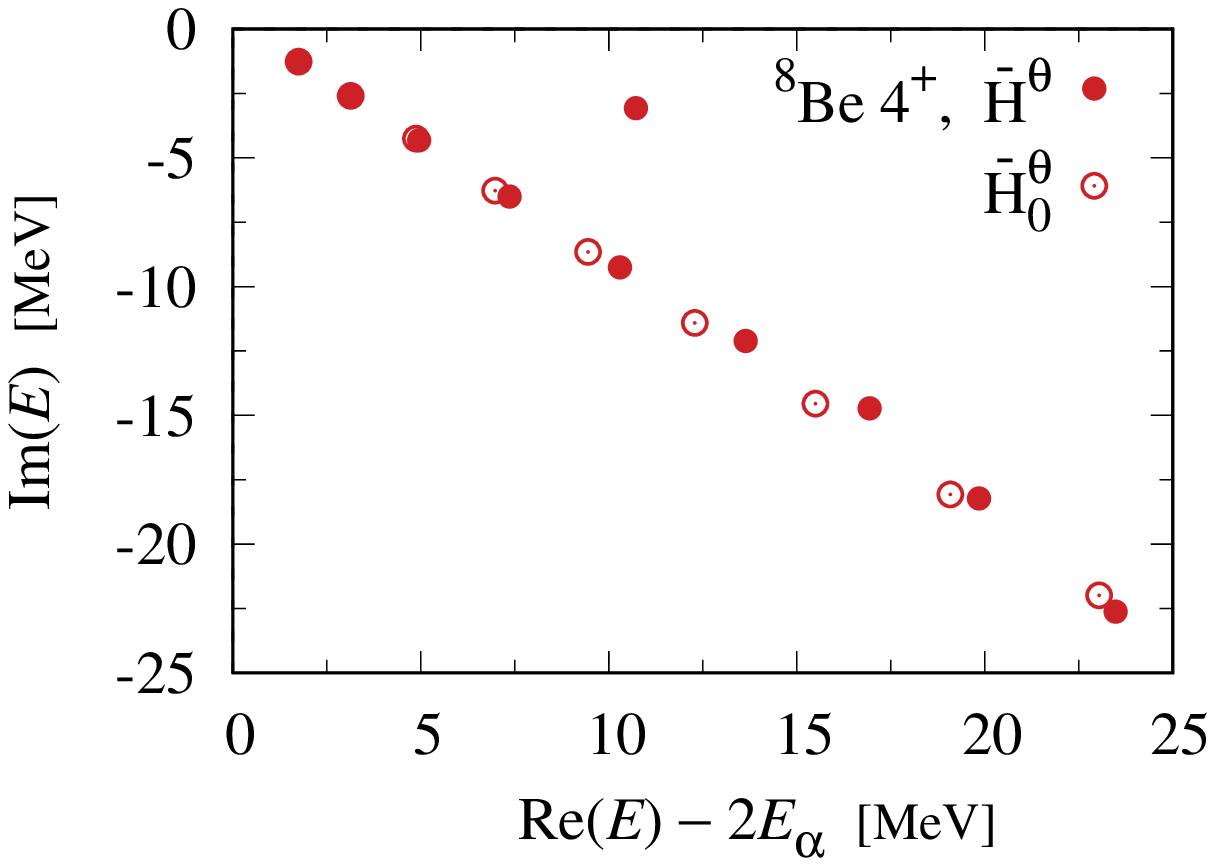}~~~~~~
\includegraphics[width=7.5cm,clip]{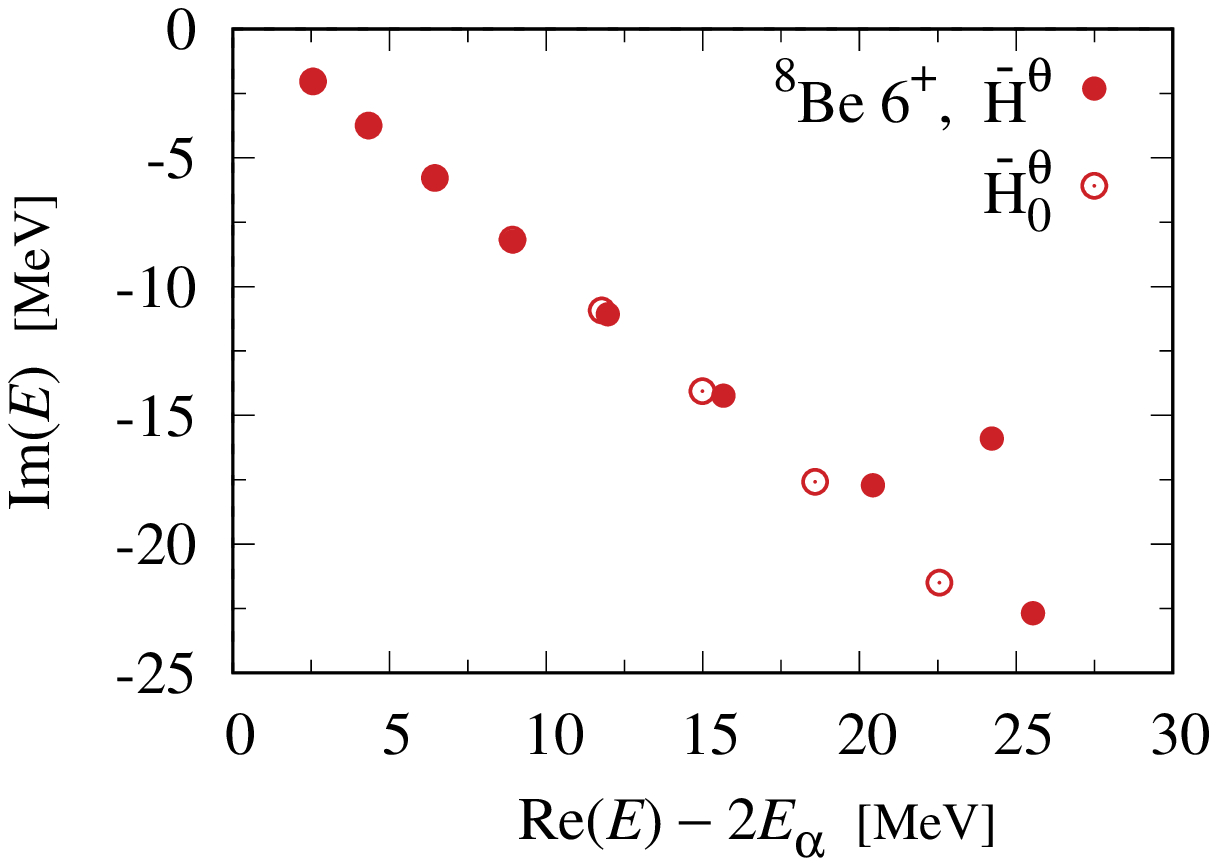}
\caption{Energy eigenvalues of $^8$Be ($0^+$, $2^+$, $4^+$, and $6^+$) for the full Hamiltonian $\bar H^\theta$ (solid symbols) and the asymptotic one $\bar H_0^\theta$ (open symbols) in the complex energy plane with $\theta=27^\circ$,
  measured from the $\alpha$+$\alpha$ threshold energy.}
\label{fig:ene_8Be_full}
\end{figure*}

We solve the complex-scaled eigenvalue problem in Eq.~(\ref{eq:eigen}) for 2$\alpha$ of $^8$Be ($0^+$, $2^+$, $4^+$, and $6^+$).
In Fig. \ref{fig:ene_8Be_full}, we show the energy eigenvalues $\{E_n^\theta\}$ of four spin states in the complex energy plane with solid symbols.
We set the scaling angle $\theta=27^\circ$, which gives the stable solutions of the energy eigenvalues of resonances and the level density.
The continuum states are discretized almost in a straight line and we find one resonance in each spin state, which deviates from the continuum states.
For the $0^+$ resonance, its energy eigenvalue is obtained very close to the $\alpha$+$\alpha$ threshold energy.
In Table \ref{tab:energy}, we list the resonance energies and decay widths of four resonances of $^8$Be in comparison with the experimental data.
One can confirm the good correspondence between them.

\begin{table}[b]
  \caption{Resonance parameters of $^8$Be measured from the $\alpha$+$\alpha$ threshold energy in MeV.
    The experimental data are in the square brackets \cite{tilley04,nndc}.}
\label{tab:energy} 
\centering
\small
\begin{tabular}{clll}
\noalign{\hrule height 0.5pt}
~$J^\pm$~~ &~~energy~~        & &~~decay width~ \\
\noalign{\hrule height 0.5pt}                                                                       
$0^+$      &~ $~~~0.08$       & &~~$~~< 1\times 10^{-3}$ \\
           &~ [~0.0918]       & &~~[$5.57(25)\times 10^{-6}$]  \\
\hline
$2^+$      &~ $~~~2.50$       & &~~$~~1.27$      \\
           &~ [ 3.12(1)]      & &~~[1.513(15)]   \\
\hline
$4^+$      &~ $~10.72$        & &~~$~~6.14$      \\
           &~ [11.44(15)]     & &~~[$\approx$ 3.5] \\
\hline
$6^+$      &~ $~24.18$        & &~~$~31.74$     \\
           &~ [$\approx$ 28]  & &~~[$\approx$ 20] \\
\noalign{\hrule height 0.5pt}
\end{tabular}
\end{table}

We also calculate the eigenstates of the asymptotic Hamiltonian $\bar H_0^\theta$ of $2\alpha$ using the asymptotic BB basis states for level density.
We employ the same parameters of the generator coordinates as used in the calculation with the full Hamiltonian $\bar H^\theta$ and set $\theta=27^\circ$.
In Fig.~\ref{fig:ene_8Be_full}, we show the energy eigenvalues $\{E_{0,n}^\theta\}$ of four spin states with open symbols as well as $\{E_{n}^\theta\}$.


\subsection{Level density and phase shift}
\begin{figure*}[t]  
\centering
\includegraphics[width=7.5cm,clip]{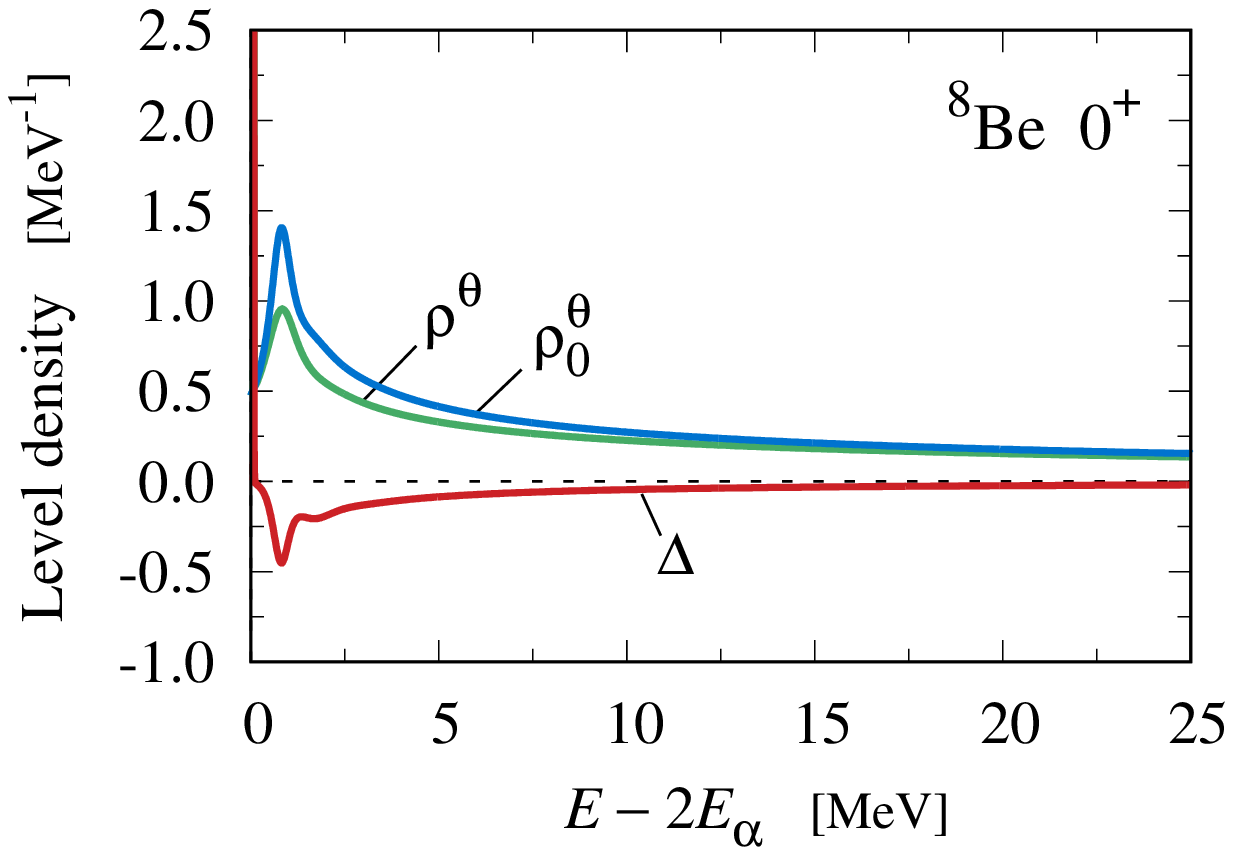}~~~~~~ 
\includegraphics[width=7.5cm,clip]{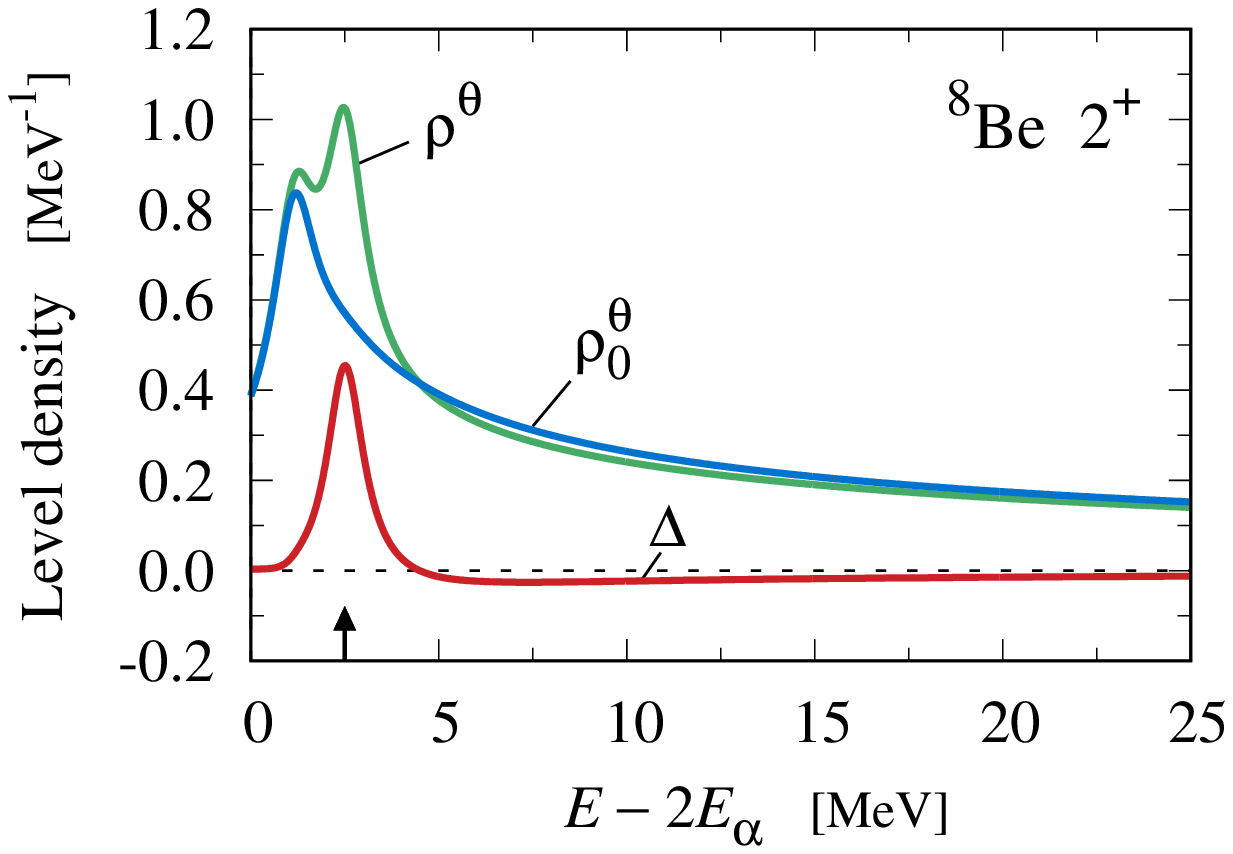}\\[0.2cm]  
\includegraphics[width=7.5cm,clip]{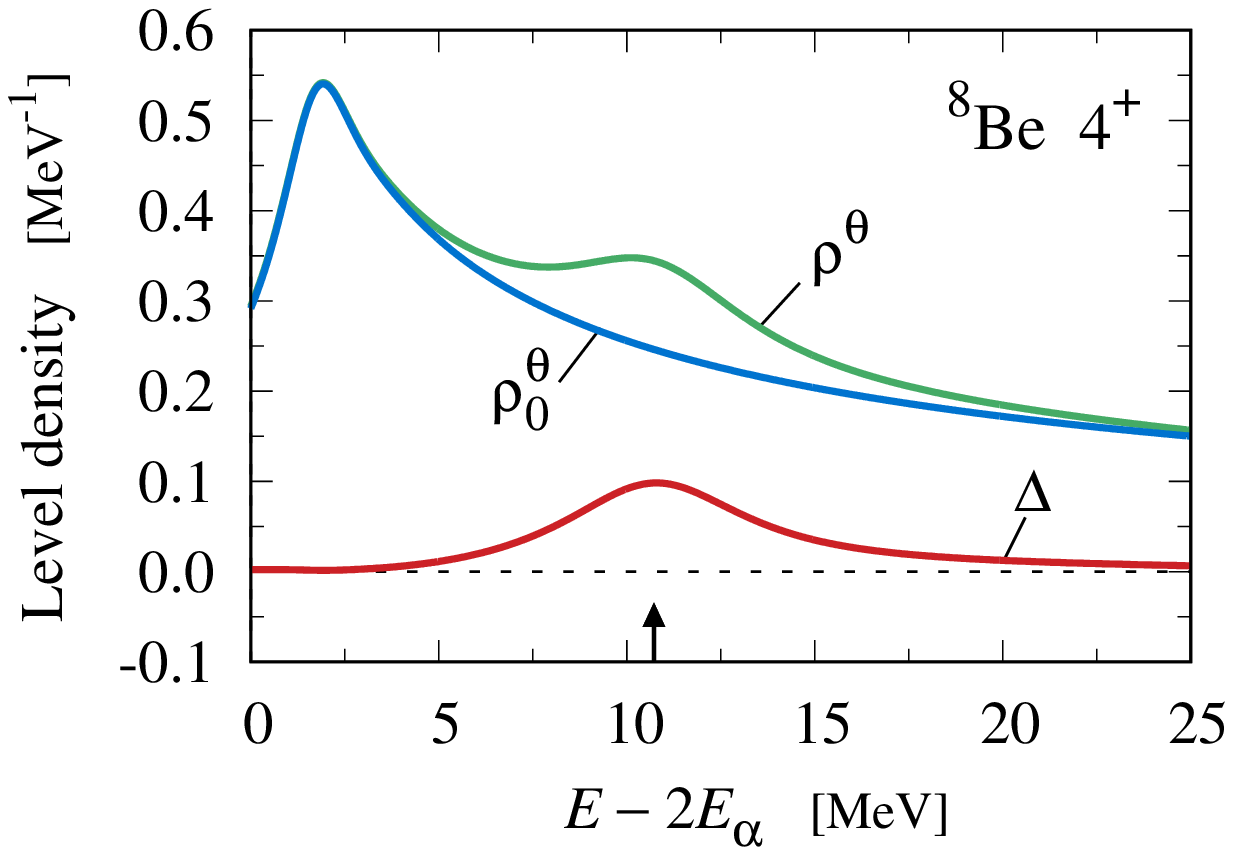}~~~~~~  
\includegraphics[width=7.5cm,clip]{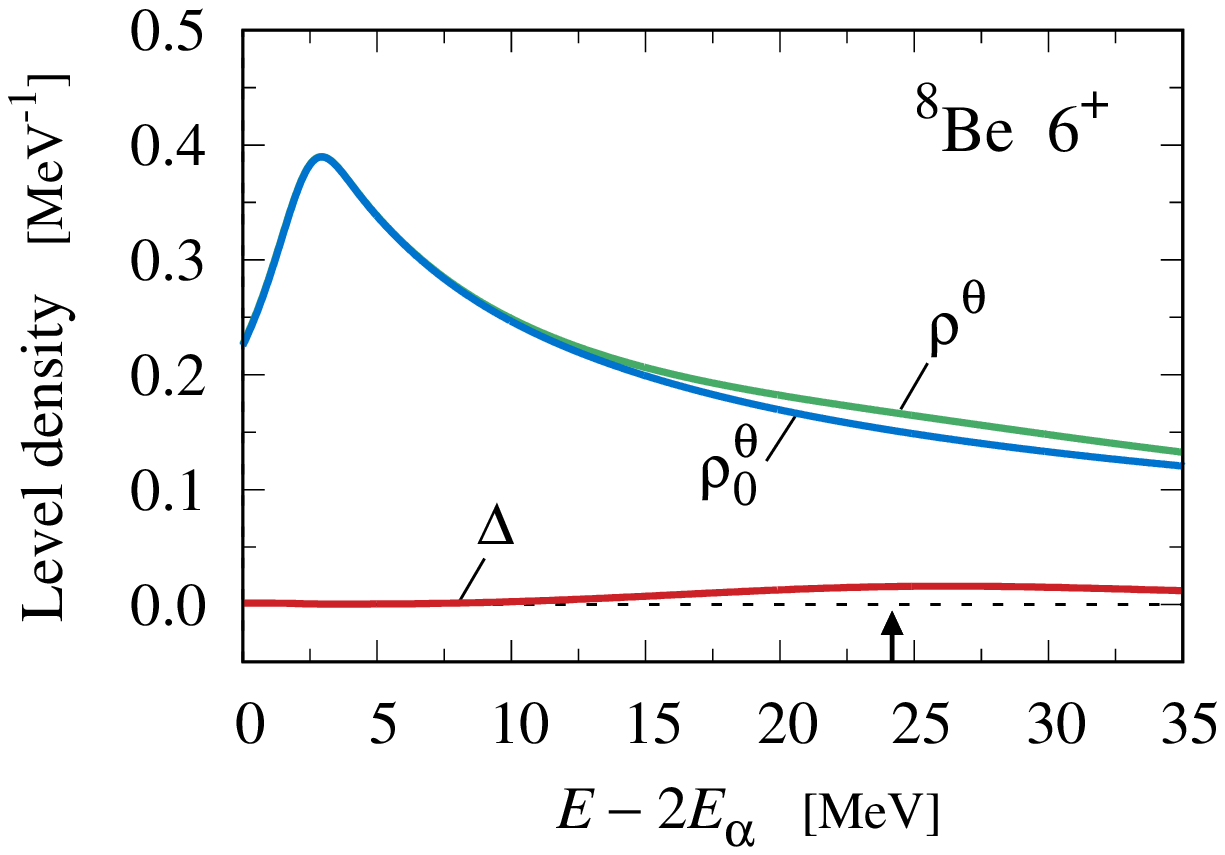}  
\caption{Two kinds of level densities, $\rho^\theta(E)$ and $\rho^\theta_0(E)$, and 
the continuum level density $\Delta(E)$ of $^8$Be ($0^+$, $2^+$, $4^+$, and $6^+$),
measured from the $\alpha$+$\alpha$ threshold energy.
The upper arrows indicate the resonance energies of $2^+$, $4^+$, and $6^+$ in Table \ref{tab:energy}.}
\label{fig:LD_8Be}
\end{figure*}

Using $\{E_n^\theta\}$ and $\{E_{0,n}^\theta\}$ of 2$\alpha$, we calculate two kinds of level densities, $\rho^\theta(E)$ and $\rho_{0}^\theta(E)$, respectively, which are used to evaluate the phase shift of the $\alpha$--$\alpha$ scattering.
In Fig.~\ref{fig:LD_8Be}, we show $\rho^\theta(E)$ and $\rho^\theta_0(E)$, the difference of which gives the continuum level density $\Delta(E)$.
For $0^+$, one confirms a very sharp peak at the zero energy in $\Delta(E)$ (red line):
there are two peaks in $\rho^\theta(E)$ (green line) at the zero energy and 1 MeV, respectively, and the peak at 1 MeV is subtracted by $\rho^\theta_0(E)$ (blue line)
as background coming from the discretized continuum states, and the remaining peak at the zero energy comes from the resonance contribution.
For $2^+$, there are two peaks in $\rho^\theta(E)$ (green line), and the one at lower energy is entirely subtracted by $\rho^\theta_0(E)$ (blue line) and
the higher peak at 2.5 MeV remains in $\Delta(E)$ (red line).
The peak in $\Delta(E)$ represents a resonance effect and the subtracted peak comes from the discretized continuum states.
For $4^+$, similarly to $2^+$, the lower-energy peak in $\rho^\theta(E)$ is subtracted by $\rho^\theta_0(E)$, and
the peak at around 11 MeV remains in $\Delta(E)$ and represents a resonance effect.
For $6^+$, $\rho^\theta(E)$ and $\rho_0^\theta(E)$ are very similar and then $\Delta(E)$ shows a very broad and small peak structure, the peak energy of which agrees with the resonance energy of 24 MeV.

For four spin states, the peak energies in $\Delta(E)$ agree with the resonance energies shown in Table \ref{tab:energy} (arrows in Fig.~\ref{fig:LD_8Be}).
Namely, from Eq.~(\ref{eq:CLD}), the energy at a maximum derivative in the phase shift fairly indicates the resonance energy.
One can discuss the existence of resonances in the distribution of $\Delta(E)$. 

\begin{figure}[t]  
\centering
\includegraphics[width=8.3cm,clip]{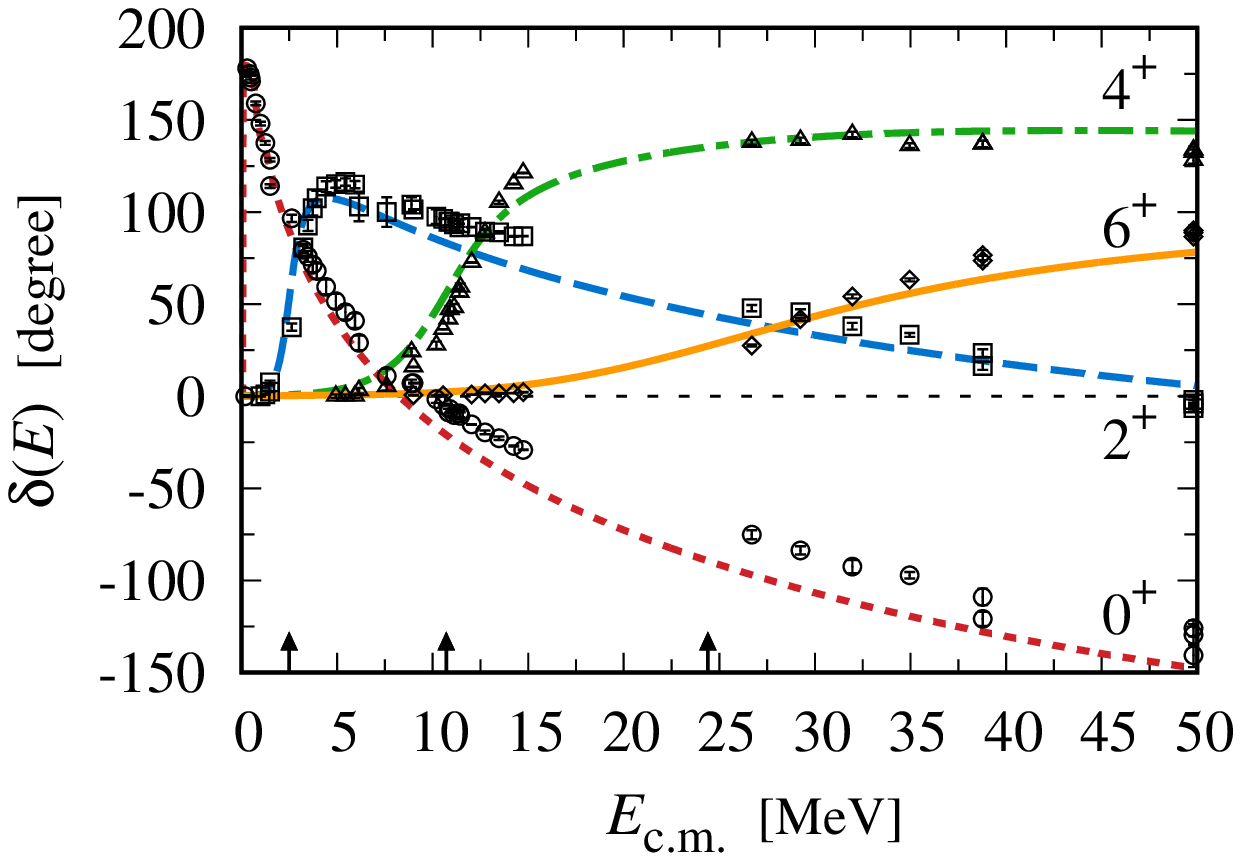}
\caption{Phase shifts of the $\alpha$--$\alpha$ scattering ($0^+$, $2^+$, $4^+$, and $6^+$) in the center-of-mass frame.
  The open symbols indicate the experimental data \cite{nilson58}.
  The upper arrows indicate the resonance energies of $2^+$, $4^+$, and $6^+$ in Table \ref{tab:energy}.}
\label{fig:PH_8Be}
\end{figure}

Finally, we evaluate the phase shift of the $\alpha$--$\alpha$ scattering by integrating $\Delta(E)$ over energy in Eq. (\ref{eq:phase}).
In Fig.~\ref{fig:PH_8Be}, we show the phase shifts of the four spin states, where we put the arrows at the resonance energies of $2^+$, $4^+$, and $6^+$.
The resulting phase shifts are consistent with the RGM calculation with $R$-matrix \cite{arai06} and also fairly reproduce the experimental data.
One can apply the complex-scaled generator coordinate method to the scattering problem between various nuclear cluster systems.

\section{Summary}\label{sec:summary}
The complex scaling is a useful method to investigate the resonances in many-body quantum systems in various physics fields.
In this paper, we presented a consistent construction of the formulation of the complex-scaled generator coordinate method for the microscopic cluster model of nuclei,
in which only the relative motions between clusters are transformed.
We applied the present framework to calculate the level density in a microscopic way, which connects to the scattering matrix.

In the generator coordinate method of the Bloch--Brink $\alpha$ cluster model,
the complex scaling is applicable by transforming the generator coordinates of each $\alpha$ cluster.
We derive the equivalence of this method and the transformation of the relative motions between clusters,
while the internal wave function of the $\alpha$ cluster is not transformed.
This framework is extendable to multicluster systems such as 3$\alpha$ and the addition of valence nucleons,
imposing a condition on the cluster wave function: each cluster does not involve the internal generator coordinate.
The harmonic oscillator shell model wave function is applicable for clusters as in $^{16}$O.
The formulation is desired to be developed to treat various types of the cluster wave functions of nuclei,
such as the $\alpha$ condensate wave function \cite{zhou20}, the antisymmetrized molecular dynamics (AMD) \cite{kanada03,takatsu23}, 
the tensor-optimized AMD \cite{myo17,myo22}, and the high-momentum AMD \cite{lyu20,isaka22},
the latter two of which can treat the bare nuclear interaction.

In this paper, we showed the reliability of the method by solving the $\alpha$--$\alpha$ scattering problem of $^8$Be.
We obtain the resonant and nonresonant continuum solutions in the complex scaling.
Using them in the Green's function, we calculate the level density and evaluate the scattering phase shifts.
We expect further applications of the present framework to the nuclear clustering phenomena involving many-body resonances
because many-body unbound states can be treated in the complex scaling.

\section*{Acknowledgments}
We would like to thank Prof. Hiroshi Toki, Prof. Hisashi Horiuchi, and Prof. Kiyoshi Kat\=o for useful discussions and comments.
This work was supported by JSPS KAKENHI Grants No. JP18K03660 and No. JP22K03643.

\section*{References}
\def\JL#1#2#3#4{ {{\rm #1}} \textbf{#2}, #4 (#3)}  
\nc{\PR}[3]     {\JL{Phys. Rev.}{#1}{#2}{#3}}
\nc{\PRC}[3]    {\JL{Phys. Rev.~C}{#1}{#2}{#3}}
\nc{\PRA}[3]    {\JL{Phys. Rev.~A}{#1}{#2}{#3}}
\nc{\PRL}[3]    {\JL{Phys. Rev. Lett.}{#1}{#2}{#3}}
\nc{\NP}[3]     {\JL{Nucl. Phys.}{#1}{#2}{#3}}
\nc{\NPA}[3]    {\JL{Nucl. Phys. A}{#1}{#2}{#3}}
\nc{\PL}[3]     {\JL{Phys. Lett.}{#1}{#2}{#3}}
\nc{\PLB}[3]    {\JL{Phys. Lett.~B}{#1}{#2}{#3}}
\nc{\PTP}[3]    {\JL{Prog. Theor. Phys.}{#1}{#2}{#3}}
\nc{\PTPS}[3]   {\JL{Prog. Theor. Phys. Suppl.}{#1}{#2}{#3}}
\nc{\PTEP}[3]   {\JL{Prog. Theor. Exp. Phys.}{#1}{#2}{#3}}
\nc{\PRep}[3]   {\JL{Phys. Rep.}{#1}{#2}{#3}}
\nc{\PPNP}[3]   {\JL{Prog.\ Part.\ Nucl.\ Phys.}{#1}{#2}{#3}}
\nc{\JPG}[3]     {\JL{J. of Phys. G}{#1}{#2}{#3}}
\nc{\andvol}[3] {{\it ibid.}\JL{}{#1}{#2}{#3}}

\end{document}